\documentclass[notitlepage]{revtex4-1}

\usepackage[utf8]{inputenc}
\usepackage[colorlinks=true,citecolor=blue,linkcolor=blue]{hyperref}
\usepackage[normalem]{ulem}
\usepackage{url}
\usepackage{graphicx,wrapfig,float,slashed,cancel}
\usepackage{amsmath,amssymb,epsfig,graphicx,xcolor,stmaryrd}
\usepackage{bm}
\usepackage{enumitem}
\usepackage{multirow}
\usepackage{soul}
\usepackage[utf8]{inputenc}
\usepackage{xcolor}
\usepackage{empheq}
\usepackage[many]{tcolorbox}

\definecolor{darkblue}{RGB}{1, 90, 173}

\begin{document}


\title{Investigations of $\Lambda$ states with spin-parity $\frac{3}{2}^{\pm}$ }

\author{K.~Azizi}
\email{ kazem.azizi@ut.ac.ir}
\thanks{Corresponding author}
\affiliation{Department of Physics, University of Tehran, North Karegar Avenue, Tehran
14395-547, Iran}
\affiliation{Department of Physics, Do\v{g}u\c{s} University, Dudullu-\"{U}mraniye, 34775
Istanbul, Turkey}
\affiliation{School of Particles and Accelerators, Institute for Research in Fundamental Sciences (IPM) P.O. Box 19395-5531, Tehran, Iran}
\author{Y.~Sarac}
\email{yasemin.sarac@atilim.edu.tr}
\affiliation{Electrical and Electronics Engineering Department,
Atilim University, 06836 Ankara, Turkey}
\author{H.~Sundu}
\email{ hayriyesundu.pamuk@medeniyet.edu.tr}
\affiliation{Department of Physics Engineering, Istanbul Medeniyet University, 34700 Istanbul, Turkey}

\date{\today}

\preprint{}

\begin{abstract}

The present study provides spectroscopic investigations of spin-$\frac{3}{2}$ $\Lambda$ baryons with both positive and negative parities. The analysis mainly focuses on three states, namely $1P$, $2P$, and $2S$, and corresponding masses are calculated using the QCD sum rule method. To implement the method, we apply two types of interpolating currents with octet and singlet quantum numbers and compare the corresponding results with the reported masses of experimentally observed states. From the comparisons, it is extracted that the results of interpolating current with octet quantum numbers are in good agreement with the experimentally measured masses. The masses obtained with this interpolating current are $m=1513.64\pm 8.76$~MeV for $1P$ state with $J^P=\frac{3}{2}^-$, $m'=1687.91\pm 0.31$~MeV for $2P$ state with $J^P=\frac{3}{2}^-$ and $\tilde{m}=1882.37 \pm 11.95$~MeV for $2S$ state with $J^P=\frac{3}{2}^+$ and they are consistent with the experimental masses of $\Lambda(1520)$, $\Lambda(1690)$ and $\Lambda(1890)$, respectively, which confirm their spin-parity quantum numbers. Besides, we calculate the corresponding current coupling constants, which are utilized as inputs in the calculations of different form factors defining the widths of the states under study.

\end{abstract}


\maketitle

\renewcommand{\thefootnote}{\#\arabic{footnote}}
\setcounter{footnote}{0}
\section{\label{sec:level1}Introduction}\label{intro} 

The advances in experimental research areas have brought many new observations of both conventional and nonconventional hadrons. As a result of the improved experimental techniques and analyses, the excited states of light and heavy baryons have been observed with progressively higher confidence levels, and our understanding of the strong interaction has been enhanced via the investigations conducted to comprehend the properties of these observed states. The discrepancy between the numbers of the observed excited nucleon and other resonance states and the expectation of the quark model keeps the subject hot and collects interest in their investigation. Consequently, comprehending the spectroscopic characteristics, substructures, and interactions of the present or newly observed states is crucial for gaining a deeper understanding of both the strong interaction and the absent resonances.

The studies on the hyperon resonances and their excited states are of importance to enhance our understanding of these states. The need for the investigation of these states is augmented by our poor knowledge of their properties compared to those of non-strange baryons such as $N$ and $\Delta$ resonances. The presence of $s$ quark with a mass heavier than the mass of $u$ and $d$ quarks and lighter than that of $c$ and $b$ quarks puts these states in an interesting place among the other baryons. The excited states of baryons have been among the recent focus of investigations as a result of new observations pointing out possible excited states of heavy and light baryons~\cite{Belle:2018mqs,Aaij:2017nav,Aaij:2018tnn,Aaij:2018yqz}. Positive and negative parity excitations of these baryons have been investigated in several works using different methods, such as the quark model~\cite{Capstick:2000qj,Valcarce:2005rr,NIsgur,Aznauryan:2007ja,Aznauryan:2012ec,NIsgur86,Glozman:1995fu,MGiannini,GGalata,DeSanctis:2014ria,Eichmann:2016nsu,Ramalho:2016buz,Shah:2018ont}, lattice QCD~\cite{Walker-Loud:2008rui,HadronSpectrum:2008xlg,Edwards:2011jj}, basis lattice front quantization approach~\cite{Vary:2018pmv}, covariant three-body Fadeev approach~\cite{Sanchis-Alepuz:2014sca}, using Dyson-Schwinger and  Bethe-Salpeter equations~\cite{Eichmann:2018adq} and QCD sum rules method~\cite{Jido:1996ia,Oka:1996zz,Lee:1998cx,Kondo:2005ur,Aliev:2014foa,Azizi:2015fqa,Azizi:2015jya,Azizi:2019dfh,Braun:2014wpa}.  To advance our understanding of particle physics and the nonperturbative regime of QCD, understanding the structure and interaction mechanisms of both heavy baryons and baryons with light quark content have been of interest. Looking at the light baryons in Particle Data Group (PDG)~\cite{Workman:2022ynf}, it is seen that there exist fifteen $N$ resonances, eight $\Delta$ resonances, ten $\Lambda$ resonances, eight $\Sigma$ resonances, three $\Xi$ resonances, and one $\Omega$ resonance with 4-star status. However, the quark model predicts more states in these energy regions~\cite{Capstick:2000qj}, whose reason may be the effective degrees of freedom of the model. These indicate a need for deeper investigation of the present and missing light baryon resonances to understand their nature, structure, and interactions. With these motivations, these states were investigated via different approaches~\cite{Capstick:2000qj,Aznauryan:2012ba,Aznauryan:2011qj}.

Among these light baryons are the $\Lambda$ hyperon and its excited states. These states have been investigated in the last decades via the data coming from $K^-p$ invariant mass spectra of hyperons~\cite{CrystalBall:2001uhc,Prakhov:2004ri,Prakhov:2008dc}, partial wave analyses of $K^-p$ reactions~\cite{Zhang:2013cua} and photo-production of $\Lambda(1405)$ from JLab~\cite{CLAS:2013rjt,CLAS:2014tbc}. The recent couple channel analyses resulted in more hyperons~\cite{Zhang:2013sva,Kamano:2016djv,Kamano:2015hxa,Fernandez-Ramirez:2015tfa,Sarantsev:2019xxm,Matveev:2019igl,Anisovich:2020lec,Klempt:2020bdu}. Besides experimental analyses, the spectrum of the $\Lambda $ baryons was investigated via various theoretical models, such as the quark model ~\cite{Isgur:1978xj,Koniuk:1979vy,Gromes:1982ze,Glozman:1997ag,Capstick:2000qj,Loring:2001ky,Faustov:2015eba,Santopinto:2014opa,Xu:2022sak,Menapara:2023rur,Menapara:2021dzi} and lattice QCD~\cite{Engel:2012qp,Menadue:2011pd,Gubler:2016viv}. The QCD sum rule method was applied to study the mass of $\Lambda$ baryons~\cite{Azizi:2023tmw,Singh:2006ii,Lee:2002jb}. The masses of the excited states of light strange baryons, including $\Lambda$ baryon, were studied in Ref.~\cite{Oudichhya:2022off} using the Regge phenomenology with quasi-linear Regge trajectories. In Ref.~\cite{Xu:2022sak}, it was pointed out that $\Lambda(1405)$, $\Lambda(1520)$, $\Lambda(1670)$, and $\Lambda(1690)$ are possible mixtures of the p-wave $q^2s$ states and ground $q^3\bar{s}q$ pentaquark states. In Ref.~\cite{Edwards:2012fx}, 71 $\Lambda^*$'s were predicted by lattice QCD calculations. Previous analyses with the constituent quark models provided similar predictions~\cite{NIsgur86,Capstick:2000qj}. All these put the baryon states containing strange quarks among the focus of experimental and theoretical researches for either the investigation of missing hyperon states or better understanding the observed ones. Besides, these studies improve our understanding of the nonperturbative regime of QCD. Their investigations provide us with a better understanding of their intrinsic structure and the degrees of freedom in a baryon and help us improve our understanding of confinement mechanisms as well. With this motivation, baryon states with strange quarks have been the subject of investigations in many experimental facilities around the world to investigate their spectroscopy and interection mechanisms (for some recent experimental investigations, see, for instance, the Refs.~\cite{Oudichhya:2022off,Crede:2023ncq,Yang:2023jdc,Dobbs:2022agy,Thiel:2020eqn,LHCb:2023ptw,Schonning:2020uyl}).

In this work, our main purpose is to study the $\Lambda$ states with spin $J=\frac{3}{2}$. With this aim, we make  spectrum analyses for $\Lambda$ state with this spin and investigate not only the ground state but also the corresponding first orbital and first radial excitations. Among $\Lambda$ states, our analyses with the considered spin-parity may supply relevant information about the states $\Lambda(1520)$ with spin-parity $J^P=\frac{3}{2}^-$, $\Lambda(1690)$ with spin-parity $J^P=\frac{3}{2}^-$, and $\Lambda(1890)$ with spin-parity $J^P=\frac{3}{2}^+$ which have the same spin-parity quantum numbers with the considered states in the present work. Such analyses require a nonperturbative method, and in this work, we apply the QCD sum rule method~\cite{Shifman:1978bx,Shifman:1978by,Ioffe81} as it is among the successful nonperturbative methods with predictions exhibiting good consistency with experimental observations. To apply the method, the main  ingredient is a proper interpolating current describing the considered state, which is composed of the quark fields according to the valence quark content of the state. We investigate the masses and current coupling constants with two types of interpolating currents with octet or flavor-singlet quantum numbers~\cite{Lee:2002jb}. The comparison of the results with the present experimental mass values helps us understand the structure of these states, gain information about the interaction mechanisms of the quarks in the low energy domain of the QCD, and may also contribute to future investigations of missing states.

The outline of the present work is as follows: In  next section,  the details of the QCD
sum rule calculations for the masses and current coupling constants of all considered states are given. The numerical analyses for the results are attained in Sec.~\ref{III}. Sec~\ref{IV} is devoted to the summary and conclusion.

\section{The QCD sum rule for the $\Lambda$ states}\label{II}

Among the efficient approaches for elucidating the structure and properties of a particular resonance is determining its mass by assigning a proper structure to the resonance. The comparison of the computed mass results with experimental findings significantly helps us understand their nature and substructures. With this angle, our purpose in the present work is to calculate the masses for the $\Lambda$ states with spin-$\frac{3}{2}$ and both negative and positive parities. To accomplish this task, we choose proper interpolating fields comprised of the quark fields consistent with the valence quark content and quantum numbers of the $\Lambda$ states. To this end, we choose two types of interpolating currents with octet and flavor-singlet quantum numbers given as~\cite{Lee:2002jb}: 
\begin{eqnarray}
\eta^{o}_{\mu}&=&\sqrt{\frac{1}{6}}\epsilon_{abc}[2(u^a{}^TC\sigma_{\kappa\delta}d^b)\sigma^{\kappa\delta}\gamma_{\mu}s^c+(u^a{}^TC\sigma_{\kappa\delta}s^b)\sigma^{\kappa\delta}\gamma_{\mu}d^c-(d^a{}^TC\sigma_{\kappa\delta}s^b)\sigma^{\kappa\delta}\gamma_{\mu}u^c],\nonumber\\
\eta^{s}_{\mu}&=&\sqrt{\frac{1}{3}}\epsilon_{abc}[(u^a{}^TC\sigma_{\kappa\delta}d^b)\sigma^{\kappa\delta}\gamma_{\mu}s^c-(u^a{}^TC\sigma_{\kappa\delta}s^b)\sigma^{\kappa\delta}\gamma_{\mu}d^c+(d^a{}^TC\sigma_{\kappa\delta}s^b)\sigma^{\kappa\delta}\gamma_{\mu}u^c].\label{Currents}
\end{eqnarray}
In these currents, $C$ denotes charge conjugation operator; $a,~b,~c$ indices are used to represent the color indices of the quark fields $u,~d~$ and $s$, and $o(s)$ in $\eta^{o(s)}$ denotes octet (singlet)-type current. These two currents are used in the following correlation function to obtain the mass sum rules:
\begin{equation}
\Pi_{\mu\nu}(q)=i\int d^{4}xe^{iq\cdot
x}\langle 0|\mathcal{T} \{\eta^{o(s)}_{\mu}(x)\bar{\eta}^{o(s)}_{\nu}(0)\}|0\rangle,
\label{eq:CorrFmassLam}
\end{equation}
where $\mathcal{T}$ represents the time ordering operator.

The calculation of the correlation function, Eq.~(\ref{eq:CorrFmassLam}), is performed in two ways called the hadronic and the QCD sides, respectively. The results obtained from these two sides are matched via a dispersion relation to obtain the QCD sum rules for the physical parameters under quest. These results contain various Lorentz structures, and in the matching procedure, one equates the coefficients of one of these Lorentz structures. The contributions coming from the higher states and continuum are suppressed by the application of the Borel transformation and continuum subtraction operations to both sides.   

In the hadronic side, the correlator is calculated treating the interpolating currents as creation and annihilation operators, which create or annihilate the considered states from the vacuum. To proceed in the calculation a complete set of hadronic states with same quantum numbers carried by the interpolating currents is inserted into the correlation function. Then, the integral over four-$x$ is performed giving the following result:
\begin{eqnarray}
\Pi^{\mathrm{Had}}_{\mu\nu}(q)&=& \frac{\langle 0|\eta^{o(s)}_{\mu}|\Lambda(q,s)\rangle \langle \Lambda(q,s)|\bar{\eta}^{o(s)}_{\nu}|0\rangle}{m^2-q^2}+\frac{\langle 0|\eta^{o(s)}_{\mu}|\Lambda'(q,s)\rangle \langle \Lambda'(q,s)|\bar{\eta}^{o(s)}_{\nu}|0\rangle}{m'^2-q^2}+\frac{\langle 0|\eta^{o(s)}_{\mu}|\tilde{\Lambda}(q,s)\rangle \langle \tilde{\Lambda}(q,s)|\bar{\eta}^{o(s)}_{\nu}|0\rangle}{\tilde{m}^2-q^2}\nonumber\\
&+&\cdots,
\label{eq:masshadronicside1}
\end{eqnarray}
where  the  explicitly presented terms  denote the contributions of negative parity spin-$\frac{3}{2}$ ground state as well as  its excitations with negative and positive parities, respectively, and $ \cdots $ shows  the contributions of higher resonances and continuum. $|\Lambda(q,s)\rangle$, $|\Lambda'(q,s)\rangle$ and $|\tilde{\Lambda}(q,s)\rangle$ correspond to their one-particle states with respective masses $m$, $m'$ and $\tilde{m}$. The  matrix elements in Eq.~(\ref{eq:masshadronicside1}) are defined in terms of current coupling constants and spin-vectors, $u_{\mu}$, in Rarita Schwinger representation as:
\begin{eqnarray}
\langle 0|\eta^{o(s)}_{\mu}|\Lambda(q,s)\rangle &=& \lambda \gamma_5 u_{\mu}(q,s),\nonumber\\
\langle 0|\eta^{o(s)}_{\mu}|\Lambda '(q,s)\rangle &=& \lambda ' \gamma_5 u_{\mu}(q,s),\nonumber\\
\langle 0|\eta^{o(s)}_{\mu}|\tilde{\Lambda}(q,s)\rangle &=& \tilde{\lambda} u_{\mu}(q,s).
\label{eq:matrixelement1}
\end{eqnarray}
Once these matrix elements are used in the Eq.~(\ref{eq:masshadronicside1}), the summation over spin is required to proceed, which has the following form:
\begin{eqnarray}\label{Rarita}
\sum_s  u_{\mu} (q,s)  \bar{u}_{\nu} (q,s) &= &-(\!\not\!{q} + m)\Big[g_{\mu\nu} -\frac{1}{3} \gamma_{\mu} \gamma_{\nu} - \frac{2q_{\mu}q_{\nu}}{3m^{2}} +\frac{q_{\mu}\gamma_{\nu}-q_{\nu}\gamma_{\mu}}{3m} \Big].
\end{eqnarray}
At this point we need to mention that the interpolating currents that we use couple also to spin-$\frac{1}{2}$  positive and negative parity states with their corresponding matrix elements given as:
\begin{eqnarray}
\langle 0|\eta^{o(s)}_{\mu}|\frac{1}{2}^+(q)\rangle =A_{\frac{1}{2}^+}(\gamma_{\mu}+\frac{4q_{\mu}}{m_{\frac{1}{2}^+}})\gamma_5 u(q,s),
\end{eqnarray}
and
\begin{eqnarray}
\langle 0|\eta^{o(s)}_{\mu}|\frac{1}{2}^-(q)\rangle =A_{\frac{1}{2}^-}(\gamma_{\mu}+\frac{4q_{\mu}}{m_{\frac{1}{2}^-}})u(q,s),
\end{eqnarray}
respectively. From these matrix elements, it can be seen that the coefficients of the Lorentz structures containing $\gamma_{\mu}$ and $q_{\mu}$ get contributions from spin-$\frac{1}{2}$ states also. To avoid these contributions and to select the ones only coming from the spin-$\frac{3}{2}$ states, we make a proper choice of Lorentz structure giving mere spin-$\frac{3}{2}$ states' contribution. 
Using the above  relations, the result of hadronic side becomes
\begin{eqnarray}\label{PhyssSide}
\Pi_{\mu\nu}^{\mathrm{Had}}(q)&=&\frac{\lambda^{2}}{q^{2}-m^{2}}(\!\not\!{q} - m)\Big[g_{\mu\nu} -\frac{1}{3} \gamma_{\mu} \gamma_{\nu} - \frac{2q_{\mu}q_{\nu}}{3m^{2}} -\frac{q_{\mu}\gamma_{\nu}-q_{\nu}\gamma_{\mu}}{3m} \Big]\nonumber \\
&+&\frac{\lambda'{}^{2}}{q^{2}-m'{}^{2}}(\!\not\!{q} - m')\Big[g_{\mu\nu} -\frac{1}{3} \gamma_{\mu} \gamma_{\nu} - \frac{2q_{\mu}q_{\nu}}{3m'{}^{2}} -\frac{q_{\mu}\gamma_{\nu}-q_{\nu}\gamma_{\mu}}{3m'} \Big]\nonumber\\
&+&\frac{\tilde{\lambda}^{2}}{q^{2}-\tilde{m}^{2}}(\!\not\!{q} +\tilde{m})\Big[g_{\mu\nu} -\frac{1}{3} \gamma_{\mu} \gamma_{\nu} - \frac{2q_{\mu}q_{\nu}}{3\tilde{m}^{2}} +\frac{q_{\mu}\gamma_{\nu}-q_{\nu}\gamma_{\mu}}{3\tilde{m}} \Big]+\cdots.
\end{eqnarray}
As it is seen, this expression contain many Lorentz structures, however, only $g_{\mu\nu}$ and $\!\not\!{q} g_{\mu\nu}$ structures are free from spin-$\frac{1}{2}$ pollution and give contributions only to   spin-$\frac{3}{2}$ states. In principle, as the standard application of QCD sum rule method, both of these structures  can be selected to predict the mass  of the desired states.  However, in our case, both of these structures lead to roughly the same results.  To this end, we consider the structure $g_{\mu\nu}$ in our analyses to extract the physical parameters of the states under study.
Hence, we have
\begin{eqnarray}
\Pi _{\mu \nu}^{\mathrm{Had}}(q)&=&-\frac{\lambda{}^2}{q^{2}-m^{2}}  m g_{\mu\nu}
-
\frac{\lambda'{}^2}{q^{2}-m'{}^{2}}  m' g_{\mu\nu} +\frac{\tilde{\lambda}{}^2}{q^{2}-\tilde{m}^{2}} \tilde{m} g_{\mu\nu} +\mbox{other structures}+\cdots.
\label{eq:CorFun1}
\end{eqnarray}
After the Borel transformation with respect to $-q^2$, which is applied to suppress the contribution coming from higher states and continuum, the final result becomes
\begin{eqnarray}
\mathcal{\widehat B}\Pi _{\mu \nu}^{\mathrm{Had}}(q)&=&\lambda^2 e^{-\frac{m^{2}}{M^{2}}} mg_{\mu\nu} 
+
\lambda'{}^2 e^{-\frac{m'^{2}}{M^{2}}} m' g_{\mu\nu}
-\tilde{\lambda}^2 e^{-\frac{\tilde{m}^{2}}{M^{2}}}\tilde{m} g_{\mu\nu} +\mbox{other structures}+\cdots.
\label{eq:CorFunBorel}
\end{eqnarray}

As for the QCD side of the calculation, the operator product expansion (OPE) is applied, and the correlation function is computed using the interpolating currents explicitly in Eq.~(\ref{eq:CorrFmassLam}). This gives a result in terms of quark fields, and, applying Wick's theorem, possible contractions between the quark fields are obtained. So, the results turn into the ones given in terms of the quark propagators. The corresponding result for the octet current is provided here to exemplify the form of the results:
\begin{eqnarray}
\Pi_{\mu \nu}^{\mathrm{QCD}}(q)&=&i\int d^{4}xe^{iq\cdot
x}\frac{1}{6} \epsilon_{abc}\epsilon_{a'b'c'}\sigma^{\kappa\delta}\gamma_{\mu}\Big \{ 4S_s^{cc'}(x)\gamma_{\nu}\sigma_{\kappa'\delta'}\mathrm{Tr}[S_d^{bb'}(x)\sigma_{\kappa'\delta'}\tilde{S}_u^{aa'}(x)\sigma^{\kappa\delta}]\nonumber\\
&-&2S_s^{cb'}(x)\sigma_{\kappa'\delta'}\tilde{S}_u^{aa'}(x)\sigma_{\kappa\delta}S_d^{bc'}(x)\gamma_{\nu}\sigma^{\kappa'\delta'}-2S_s^{cb'}(x)\sigma_{\kappa'\delta'}\tilde{S}_d^{ba'}(x)\sigma_{\kappa\delta}S_u^{ac'}(x)\gamma_{\nu}\sigma^{\kappa'\delta'}\nonumber\\
&-&2S_d^{cb'}(x)\sigma_{\kappa'\delta'}\tilde{S}_u^{aa'}(x)\sigma_{\kappa\delta}S_s^{bc'}(x)\gamma_{\nu}\sigma^{\kappa'\delta'}+ S_d^{cc'}(x)\gamma_{\nu}\sigma_{\kappa'\delta'}\mathrm{Tr}[S_s^{bb'}(x)\sigma_{\kappa'\delta'}\tilde{S}_u^{aa'}(x)\sigma^{\kappa\delta}]\nonumber\\
&+& S_d^{ca'}(x)\sigma_{\kappa'\delta'}\tilde{S}_s^{bb'}(x)\sigma_{\kappa\delta}S_u^{ac'}(x)\gamma_{\nu}\sigma^{\kappa'\delta'}-2S_u^{ca'}(x)\sigma_{\kappa'\delta'}\tilde{S}_d^{ab'}(x)\sigma_{\kappa\delta}S_s^{bc'}(x)\gamma_{\nu}\sigma^{\kappa'\delta'}\nonumber\\
&+&S_u^{ca'}(x)\sigma_{\kappa'\delta'}\tilde{S}_s^{bb'}(x)\sigma_{\kappa\delta}S_d^{ac'}(x)\gamma_{\nu}\sigma^{\kappa'\delta'}+S_u^{cc'}(x)\gamma_{\nu}\sigma_{\kappa'\delta'}\mathrm{Tr}[S_s^{bb'}(x)\sigma_{\kappa'\delta'}\tilde{S}_d^{aa'}(x)\sigma^{\kappa\delta}]\Big\},
\end{eqnarray}
where $\tilde{S}_q^{aa}$ represents $CS_q^{aa'T}C$. The quark propagator for light quark has the following form in $x$ space:
\begin{eqnarray}
 S_{q}^{ab}(x)&=&i\frac{x\!\!\!/}{2\pi^{2}x^{4}}\delta_{ab}-\frac{m_{q}}{4\pi^{2}x^{2}}\delta_{ab}-\frac{\langle
 \overline{q}q\rangle}{12}\Big(1-i\frac{m_{q}}{4}x\!\!\!/\Big)\delta_{ab}-\frac{x^{2}}{192}m_{0}^{2}\langle
 \overline{q}q\rangle\Big( 1-i\frac{m_{q}}{6}x\!\!\!/\Big)\delta_{ab}-\frac{ig_{s}G_{ab}^{\theta\eta}}{32\pi^{2}x^{2}}\Big[x\!\!\!/\sigma_{\theta\eta} +\sigma_{\theta\eta}x\!\!\!/ \Big]
 \nonumber\\&-&\frac{x\!\!\!/ x^{2}g_s^2}{7776}\langle
 \overline{q}q\rangle^2\delta_{ab}-\frac{x^4\langle
 \overline{q}q\rangle\langle
 g_s^2G^2\rangle}{27648}\delta_{ab}+\frac{m_q}{32\pi^2}[ln(\frac{-x^2\Lambda^2}{4})+2 \gamma_E]g_{s}G_{ab}^{\theta\eta}\sigma_{\theta\eta}.
 \label{propagator}
\end{eqnarray}  
In the light quark propagator, Eq.~(\ref{propagator}), $\gamma_E \simeq 0.577$ is the Euler constant and $\Lambda$ is the QCD scale parameter. The calculation of this side is straightforward using this propagator in the correlator. After the applications of the Fourier and Borel transformations as well as continuum subtraction, the result takes the following form:
\begin{eqnarray}
\mathcal{\widehat B}\Pi^{\mathrm{QCD}}(s_0,M^2)&=&\int_{(m_u+m_d+m_s)^2}^{s_0} e^{-\frac{s}{M^2}}\rho(s) ds+\Gamma(M^2),
\end{eqnarray}
where $\rho(s)$ and $\Gamma(M^2)$ are the results obtained considering the coefficient of the structure $g_{\mu\nu}$ as in the hadronic side and $ s_0 $ is the continuum threshold parameter. Note that we apply the quark-hadron duality assumption to omit the suppressed contributions  of the higher states and continuum from the hadronic side with their equivalent contributions from the QCD side above the threshold $ s_0 $.

The match of the coefficients of the selected structure  from both the hadronic and QCD sides gives the  following QCD sum rule:
\begin{eqnarray}
m \lambda{}^2e^{-\frac{m^2}{M^2}}+m'\lambda'{}^2e^{-\frac{m'^2}{M^2}}-\tilde{m} \tilde{\lambda}{}^2e^{-\frac{\tilde{m}^2}{M^2}}=\mathcal{\widehat B}\Pi^{\mathrm{QCD}}(s_0,M^2).\label{EqQCDsumrule}
\end{eqnarray} 

To extract masses and current coupling constants from Eq.~(\ref{EqQCDsumrule}), we consider each state one by one and follow such an approach: First, we calculate the mass for the ground state from the first term in the left-hand side of Eq.~(\ref{EqQCDsumrule}), keeping others in continuum, namely the ground state $+$ continuum frame. To get the mass of the ground state, after taking the derivative of Eq.~(\ref{EqQCDsumrule}) with respect to $-\frac{1}{M^2}$, we divide this by Eq.~(\ref{EqQCDsumrule}) itself, i.e.:
\begin{eqnarray}
m^2=\frac{\frac{d}{d(-\frac{1}{M^2})}\mathcal{\widehat B}\Pi^{\mathrm{QCD}}(s_0,M^2)}{\mathcal{\widehat B}\Pi^{\mathrm{QCD}}(s_0,M^2)}. 
\end{eqnarray}  
The mass predicted in this way is applied together with Eq.~(\ref{EqQCDsumrule}), and the current coupling constant for this state is attained as
\begin{eqnarray}
\lambda^2=e^{\frac{m^2}{M^2}}\mathcal{\widehat B}\Pi^{\mathrm{QCD}}(s_0,M^2).
\end{eqnarray}
The mass and current coupling constants for the excited states are obtained similarly: To get results for the first radial excitation, the first two terms in Eq.~(\ref{EqQCDsumrule}) are considered, and the remaining term is treated to be inside the continuum, which means ground state + first radial ($2P$) excitation + continuum frame is applied using the results for the ground state as input. And finally, the same treatment is applied for the next excited state, namely the $2S$ state; that is, ground state + first radial excitation + $2S$ + continuum frame is now taken into account to get the physical parameters for the $2S$ state. The numerical analyses of the results for these states are given in the next section.

\section{Numerical Analyses}\label{III}

The QCD sum rules, attained for masses and current coupling constants in the previous section, are numerically analyzed in this section with the necessary input parameters. Some of these parameters are given in Table~\ref{tab:Inputs}. 
\begin{table}[h!]
\begin{tabular}{|c|c|}
\hline\hline
Parameters & Values \\ \hline\hline
$m_{u}$                                    & $2.16^{+0.49}_{-0.26}~\mathrm{MeV}$ \cite{Workman:2022ynf}\\
$m_{d}$                                    & $4.67^{+0.48}_{-0.17}~\mathrm{MeV}$ \cite{Workman:2022ynf}\\
$m_{s}$                                     & $93.4^{+8.6}_{-3.4}~\mathrm{MeV}$ \cite{Workman:2022ynf}\\
$\langle \bar{q}q \rangle (1\mbox{GeV})$    & $(-0.24\pm 0.01)^3$ $\mathrm{GeV}^3$ \cite{Belyaev:1982sa}  \\
$\langle \bar{s}s \rangle $                 & $0.8\langle \bar{q}q \rangle$ \cite{Belyaev:1982sa} \\
$m_{0}^2 $                                  & $(0.8\pm0.1)$ $\mathrm{GeV}^2$ \cite{Belyaev:1982sa}\\
$\langle \overline{q}g_s\sigma Gq\rangle$   & $m_{0}^2\langle \bar{q}q \rangle$ \\
$\langle \frac{\alpha_s}{\pi} G^2 \rangle $ & $(0.012\pm0.004)$ $~\mathrm{GeV}^4 $\cite{Belyaev:1982cd}\\
\hline\hline
\end{tabular}%
\caption{Some input parameters required for the numerical analyses.}
\label{tab:Inputs}
\end{table} 
The parameters present in Table~\ref{tab:Inputs} are necessary but not the only required parameters. Besides, the results contain two more auxiliary parameters: the Borel parameter $M^2$ and the threshold parameter $s_0$. To be able to fix these parameters, the QCD sum rules have some prescriptions that are standard for the method. These are the weak dependence of the results on these auxiliary parameters, the convergence of the OPE, and pole dominance. By OPE convergence, we mean:  The perturbative part exceeds the nonperturbative contributions and  the higher the dimension of the nonperturbative operator, the lower its contribution. Pole dominance guarantees the dominant contributions of the considered resonances compared to the higher states and continuum.

Indeed, the Borel parameter  is fixed by the dominance of the considered first three resonances over the higher states and continuum, which sets its upper limit, as well as the OPE convergence, which sets its lower limit. The Borel mass intervals obtained considering these restrictions are presented in Table~\ref{tab:results}. In this table, we also present the working intervals of the threshold parameters fixed by the ground state + $2P$ excitation + $2S$ excitation + continuum approach, as explained in the previous section. First, considering the relation of the threshold parameter to the energy of the next excited state, we fix the threshold parameter by ground state + continuum frame and get the mass and corresponding current coupling constant for the ground state, whose values are also present in Table~\ref{tab:results}. These results are used as inputs in the calculations of the mass and current coupling constant for the next excited state. To make this calculation, now the ground state + $2P$ excitation + continuum frame is taken into account, and similar to the previous calculation, the interval for the new threshold parameter is determined. With this threshold parameter interval, again, the mass and current coupling constant for this excited state are obtained and presented in Table~\ref{tab:results}. Finally, the mass and current coupling constant for the next excited state are obtained from ground state + $2P$ excitation + $2S$ excitation + continuum frame, and the threshold parameter interval for this new consideration is decided in a similar way. With this new threshold interval, the results for mass and current coupling constant in this case are also obtained and given in Table~\ref{tab:results}. The errors of the results are presented in Table~\ref{tab:results}, as well, and these errors are due to the uncertainties present in the input parameters and the determination of the working intervals of the auxiliary parameters. 

\begin{figure}[h!]
\begin{center}
\includegraphics[totalheight=5cm,width=7cm]{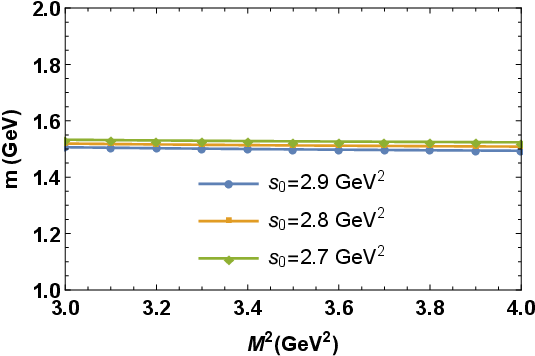}
\includegraphics[totalheight=5cm,width=7cm]{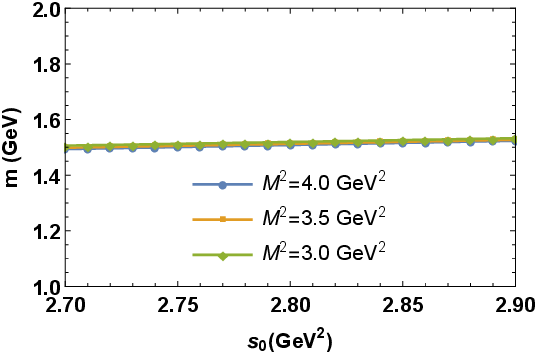}
\end{center}
\caption{\textbf{Left:} The dependence  of the mass of the ground state $\Lambda$ on  $M^2$ at different values of  $s_0$.
\textbf{Right:} The dependence of the mass of the ground state $\Lambda$ on $s_0$ at different values of   $M^2$. }
\label{gr:MassMsqS01P}
\end{figure}

\begin{figure}[h!]
\begin{center}
\includegraphics[totalheight=5cm,width=7cm]{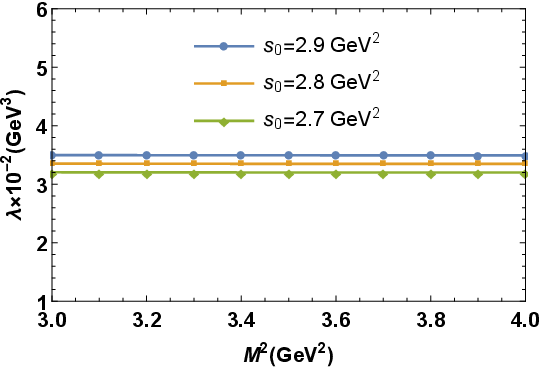}
\includegraphics[totalheight=5cm,width=7cm]{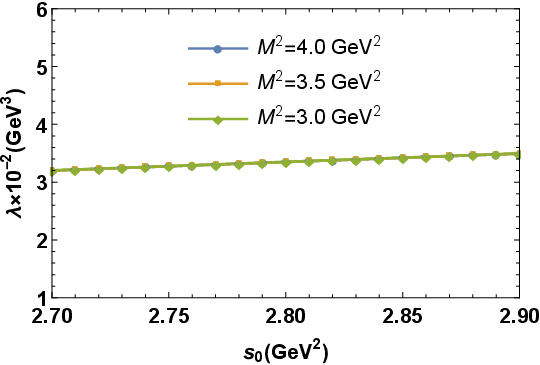}
\end{center}
\caption{\textbf{Left:} The dependence of the current coupling constant of the ground state $\Lambda$ on  $M^2$ at different values of  $s_0$.
\textbf{Right:} The dependence of the current coupling constant of the ground state $\Lambda$ on  $s_0$ at different values of  $M^2$. }
\label{gr:lamMsqS01P}
\end{figure}

\begin{figure}[h!]
\begin{center}
\includegraphics[totalheight=5cm,width=7cm]{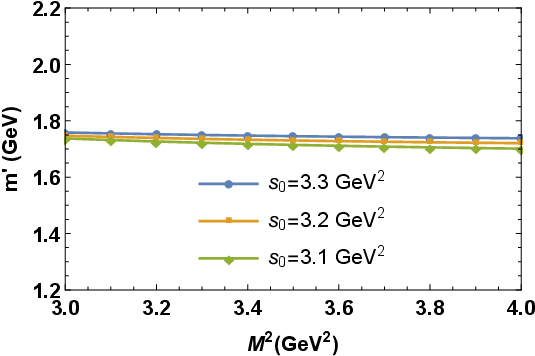}
\includegraphics[totalheight=5cm,width=7cm]{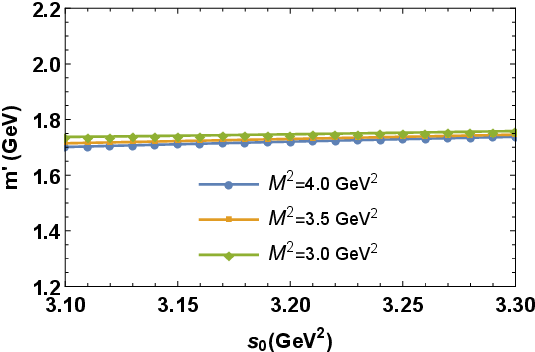}
\end{center}
\caption{\textbf{Left:} The dependence  of the mass of the $(2P)$ exited state of $\Lambda$ on  $M^2$ at different values of  $s_0$.
\textbf{Right:} The dependence of the mass of the $(2P)$ exited state of $\Lambda$ on  $s_0$ at different values of   $M^2$. }
\label{gr:MassMsqS02P}
\end{figure}

\begin{figure}[h!]
\begin{center}
\includegraphics[totalheight=5cm,width=7cm]{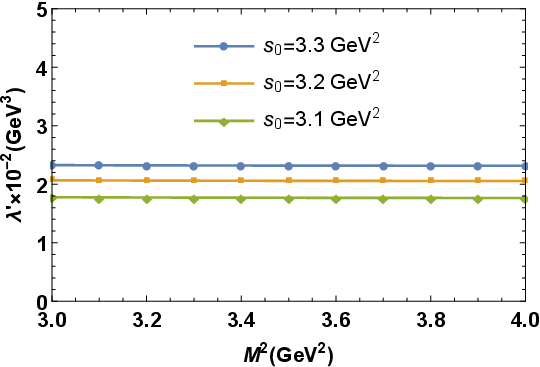}
\includegraphics[totalheight=5cm,width=7cm]{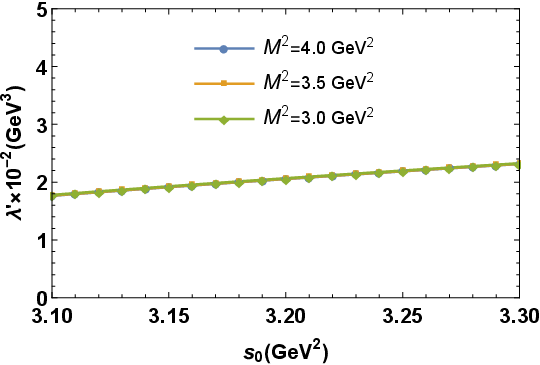}
\end{center}
\caption{\textbf{Left:} The dependence of the current coupling constant of the $(2P)$ exited state of $\Lambda$ on  $M^2$ at different values of $s_0$.
\textbf{Right:} The dependence of the current coupling constant of the $(2P)$ exited state of $\Lambda$ on  $s_0$ at different values of  $M^2$. }
\label{gr:lamMsqS02P}
\end{figure}

\begin{figure}[h!]
\begin{center}
\includegraphics[totalheight=5cm,width=7cm]{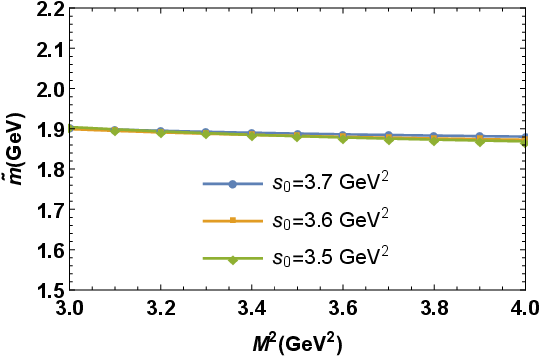}
\includegraphics[totalheight=5cm,width=7cm]{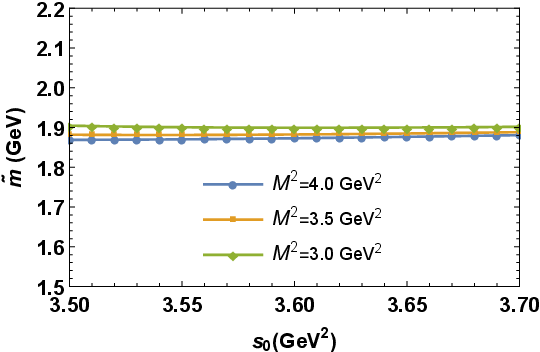}
\end{center}
\caption{\textbf{Left:} The dependence  of the mass of the $(2S)$ exited state of $\Lambda$ on  $M^2$ at different values of  $s_0$.
\textbf{Right:} The dependence of the mass of the $(2S)$ exited state of $\Lambda$ on  $s_0$ at different values of  $M^2$. }
\label{gr:MassMsqS02S}
\end{figure}

\begin{figure}[h!]
\begin{center}
\includegraphics[totalheight=5cm,width=7cm]{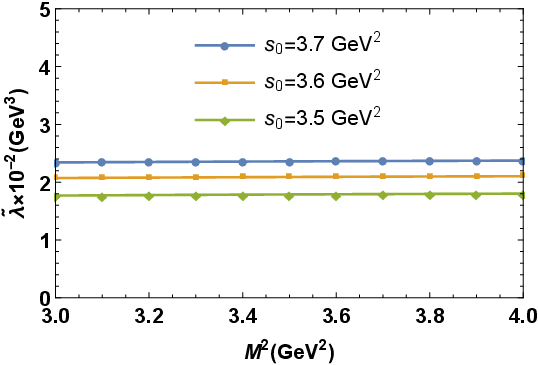}
\includegraphics[totalheight=5cm,width=7cm]{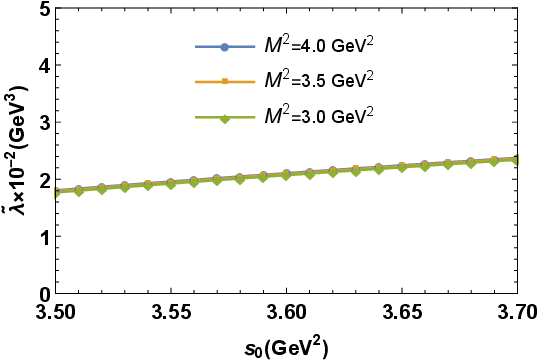}
\end{center}
\caption{\textbf{Left:} The dependence of the current coupling constant of the $(2S)$ exited state of $\Lambda$ on  $M^2$ at different values of  $s_0$.
\textbf{Right:} The dependence of the current coupling constant of the $(2S)$ exited state of $\Lambda$ on  $s_0$ at different values of  $M^2$. }
\label{gr:lamMsqS02S}
\end{figure}

It is instructive to find the first three resonance' contribution (FTRC) at average values of the Borel parameter and continuum threshold.  It is defined as
\begin{eqnarray}
\mathrm{FTRC}=\frac{\mathcal{\widehat B}\Pi^{\mathrm{QCD}}(s_0,M^2)}{\mathcal{\widehat B}\Pi^{\mathrm{QCD}}(\infty,M^2)}.
\end{eqnarray}
By using the average values of the auxiliary parameters, we find  $\mathrm{FTRC}=0.91$ indicating that the main contribution in the QCD side comes from the first three resonances and the higher states contribute with only $ 9\% $. This well-satisfies the corresponding requirement of the method.
To depict the behavior and stability of the results in the working intervals of the auxiliary parameters, we present the figures~\ref{gr:MassMsqS01P}, \ref{gr:lamMsqS01P}, \ref{gr:MassMsqS02P}, \ref{gr:lamMsqS02P}, \ref{gr:MassMsqS02S}, and \ref{gr:lamMsqS02S} for the masses and current coupling constants of all the considered states. From these figures, we see good stability of the results with respect to the auxiliary parameters. The mild variations of the results with respect to  the variations of these  parameters, seen in these figures,  are reflected in the errors of the predictions presented in Table~\ref{tab:results} as mentioned before.
\begin{table}[]
\begin{tabular}{|ccccc|}
\hline
\multicolumn{5}{|c|}{Results for octet current}                                                                                                                                                                                                                                                       \\ \hline\hline
\multicolumn{1}{|c|}{$\Lambda$ state}                  & \multicolumn{1}{c|}{$M^2~(\mathrm{GeV}^2)$} & \multicolumn{1}{c|}{$s_0~(\mathrm{GeV}^2)$} & \multicolumn{1}{c|}{Mass$~(\mathrm{MeV})$} & \begin{tabular}[c]{@{}c@{}}Current Coupling  Constant$ ~(\mathrm{GeV}^3)$\end{tabular} \\ \hline\hline
\multicolumn{1}{|c|}{$\Lambda(J^P=\frac{3}{2}^{-})(1P)$}                       & \multicolumn{1}{c|}{$3.0-4.0$}                  & \multicolumn{1}{c|}{$2.7-2.9$}              & \multicolumn{1}{c|}{$1513.64\pm8.76$}     & $(3.35\pm0.15)\times10^{-2}$                                                              \\ \hline
\multicolumn{1}{|c|}{$\Lambda(J^P=\frac{3}{2}^{-})(2P)$}                       & \multicolumn{1}{c|}{$3.0-4.0$}                  & \multicolumn{1}{c|}{$3.1-3.3$}              & \multicolumn{1}{c|}{$1687.91\pm0.31$}     & $(2.05\pm0.27)\times10^{-2}$                                                              \\ \hline
\multicolumn{1}{|c|}{$\Lambda(J^P=\frac{3}{2}^{+})(2S)$}                       & \multicolumn{1}{c|}{$3.0-4.0$}                  & \multicolumn{1}{c|}{$3.5-3.7$}              & \multicolumn{1}{c|}{$1882.37\pm11.95$}    & $(2.08\pm0.30)\times10^{-2}$                                                              \\ \hline
\multicolumn{5}{|c|}{Results for singlet current}                                                                                                                                                                                                                                                     \\ \hline\hline
\multicolumn{1}{|c|}{$\Lambda$ state} & \multicolumn{1}{c|}{$M^2~(\mathrm{GeV}^2)$} & \multicolumn{1}{c|}{$s_0~(\mathrm{GeV}^2)$} & \multicolumn{1}{c|}{Mass$~(\mathrm{MeV})$} & \begin{tabular}[c]{@{}c@{}}Current Coupling  Constant $~(\mathrm{GeV}^3)$\end{tabular} \\ \hline\hline
\multicolumn{1}{|c|}{$\Lambda(J^P=\frac{3}{2}^{-})(1P)$}                       & \multicolumn{1}{c|}{$3.0-4.0$}                  & \multicolumn{1}{c|}{$2.7-2.9$}              & \multicolumn{1}{c|}{$1470.44\pm16.08$}    & $(4.85\pm0.20)\times10^{-2}$                                                              \\ \hline
\multicolumn{1}{|c|}{$\Lambda(J^P=\frac{3}{2}^{-})(2P)$}                       & \multicolumn{1}{c|}{$3.0-4.0$}                  & \multicolumn{1}{c|}{$3.1-3.3$}              & \multicolumn{1}{c|}{$1732.56\pm11.55$}    & $(2.90\pm0.39)\times10^{-2}$                                                              \\ \hline
\multicolumn{1}{|c|}{$\Lambda(J^P=\frac{3}{2}^{+})(2S)$}                       & \multicolumn{1}{c|}{$3.0-4.0$}                  & \multicolumn{1}{c|}{$3.5-3.7$}              & \multicolumn{1}{c|}{$1843.47\pm11.21$}    & $(3.00\pm0.41)\times10^{-2}$                                                              \\ \hline
\end{tabular}
\caption{The working windows of the Borel masses and threshold values and the mass and current coupling constant results obtained using the octet and singlet currents for the $\Lambda$ states with spin-$\frac{3}{2}$ and different parities}
\label{tab:results}
\end{table}

\section{Summary and conclusion}\label{IV}

In the present work, among the light hyperon states, we focused on the $\Lambda$ state with spin-parity quantum numbers $J=\frac{3}{2}^-$ and its excitations with negative and positive parities. In the analyses, we especially targeted to account for the $\Lambda(1520)$ through its spectroscopic property, together with to see whether any experimentally observed state possessing the identical quantum numbers can be explained via the results attained for the excited states. To analyze these states, we applied the QCD sum rule method with two possible currents that may define these states with octet and singlet quantum  numbers. From the analyses, we got the mass values of the three lowest-lying states. The results were obtained using the current with octet quantum numbers as: $m=1513.64\pm 8.76$~MeV for the $1P$ state with $J^P=\frac{3}{2}^-$, $m'=1687.91\pm 0.31$~MeV for the $2P$ state with $J^P=\frac{3}{2}^-$ and $\tilde{m}=1882.37 \pm 11.95$~MeV for the $2S$ state with $J^P=\frac{3}{2}^+$. The mass values using current with singlet quantum numbers were obtained as:  $m= 1470.44\pm 16.08$~MeV for the $1P$ state with $J^P=\frac{3}{2}^-$, $m'=1732.56 \pm 11.55$~MeV for the $2P$ state with $J^P=\frac{3}{2}^-$ and $\tilde{m}= 1843.47 \pm 11.21$~MeV for the $2S$ state with $J^P=\frac{3}{2}^+$. In other works present in the literature, the masses for the lowest-lying $\Lambda$ states with spin-$\frac{3}{2}$ consistent with the ones we search for in the present work were predicted. For completeness and comparison with our predictions, we present the results of some of these works here. In Ref.~\cite{NIsgur}, the masses for the lowest-lying $P$-wave $\Lambda$ states with spin-$\frac{3}{2}$ were given as $m=1490$~MeV and $m=1690$~MeV. In Ref.~\cite{NIsgur86}, the masses for the same negative parity lowest-lying states were obtained as $m=1545$~MeV and $m=1645$~MeV for $\frac{3}{2}^-$ states, and the mass was obtained as $m=1900$~MeV for $J^P=\frac{3}{2}^+$ positive parity state. The spectral values were predicted in Ref.~\cite{Glozman:1995fu} as $m=1498$~MeV and $m=1629$~MeV for $\frac{3}{2}^-$ states and $m=1855$~MeV for $J^P=\frac{3}{2}^+$ state. In Ref.~\cite{Isgur:1978xj} the masses were given as $m=1490$~MeV, and $m=1690$~MeV for $\frac{3}{2}^-$ states. In Ref.~\cite{Loring:2001ky} the masses for spin-parity $\frac{3}{2}^{-}$ states were attained as $m=1508$~MeV, and $m=1662$~MeV, and for spin-parity $\frac{3}{2}^+$ as $m=1823$~MeV. The masses were given as $m=1549$~MeV and $m=1693$~MeV for $\frac{3}{2}^-$ states and $m=1854$~MeV for $J^P=\frac{3}{2}^+$ state in Ref.~\citep{Faustov:2015eba}. In Ref.~\cite{Menapara:2021dzi} the mass predictions for $J^P=3/2^-$ state corresponding to $\Lambda(1520)$ were calculated as $m=1534$~MeV and $m=1544$~MeV, the masses for $J^P=3/2^-$ state given for $\Lambda(1690)$ were predicted to be $m=1819$~MeV and $m=1841$~MeV and the masses for $J^P=3/2^+$ state for $\Lambda(1890)$ were predicted as $m=1769$~MeV and $m=1789$~MeV. The mass values were given in Ref.~\cite{Santopinto:2014opa} as $m=1431$~MeV, $m=1650$~MeV, and $m=1896$~MeV for $\Lambda(1520)$, $\Lambda(1690)$, and $\Lambda(1890)$ spin-$\frac{3}{2}$ states, respectively. The mass for the $1P$ spin-$\frac{3}{2}$ state was given in Ref.~\cite{Oudichhya:2022off} as $m=1551.23 \pm0.43$~MeV. 

Compared to the experimental findings, our predictions obtained from the current with octet quantum numbers are consistent with the masses of $\Lambda(1520)$, $\Lambda(1690)$ and $\Lambda(1890)$ states possessing mass values and quantum numbers $m_{\Lambda(1520)}\approx 1519$~MeV and $J^P=\frac{3}{2}^-$, $m_{\Lambda(1690)}\approx 1690$~MeV and $J^P=\frac{3}{2}^-$, and $m_{\Lambda(1890)}\approx 1890$~MeV and $J^P=\frac{3}{2}^+$, respectively~\cite{Workman:2022ynf}. Given the consistency of these results with experimental observations, it is also pertinent to compare them with the predictions of other studies. The mass result obtained using the interpolating current with octet quantum number in the present work for the $2P$ state is consistent with that of Ref.~\cite{NIsgur} given for the excited state, however, the $1P$ state is larger than their corresponding result. Compared to the results of Ref.~\cite{NIsgur86}, our value for the $1P$ state is smaller, and the $2P$ state is larger than their corresponding predictions. In Ref.~\cite{Glozman:1995fu}, the reported result for $\Lambda(1520)$ state is close to our prediction, but their predictions for $\Lambda(1690)$ and $\Lambda(1890)$ are smaller than ours. The results given in Ref.~\cite{Isgur:1978xj} are close to those of the present work obtained for $2P$ and $2S$ cases, and their result corresponding to the $1P$ case is smaller than our prediction. While the result we obtained for $\Lambda(1520)$ is consistent with that of Ref.~\cite{Loring:2001ky}, the results obtained for the other two states in that work are smaller than the predictions of the present work. The predictions for the mass of negative parity states in Ref.~\cite{Faustov:2015eba} are larger, on the other hand, the positive parity state is smaller than our corresponding predictions. In Ref.~\citep{Menapara:2021dzi}, the mass predictions given for $\Lambda(1520)$ and $\Lambda(1690)$ states are larger, and that for $\Lambda(1890)$ is smaller than ours. The reported masses for $\Lambda(1520)$ and $\Lambda(1690)$ in Ref.~\cite{Santopinto:2014opa} are smaller than our corresponding results, while the result given for $\Lambda(1890)$ is in agreement within the errors with our prediction. To provide a more explicit illustration of the aforementioned comparison, we present the Table~\ref{tab:differentworks} listing the results from other methods, experiment, and the results obtained from our analyses.

\begin{table}[]
\begin{tabular}{|c|c|c|c|c|c|c|c|c|c|c|c|c|c|}
\hline
                               &  \multicolumn{1}{c|}{\begin{tabular}[c]{@{}c@{}}Present Work\\ (Results of \\octet current)\end{tabular}} &\multicolumn{1}{c|}{\begin{tabular}[c]{@{}c@{}}Present Work\\ (Results of \\ singlet current )\end{tabular}}& \cite{NIsgur} & \cite{NIsgur86} &\cite{Glozman:1995fu}  &\cite{Isgur:1978xj}  &\cite{Loring:2001ky}  &\cite{Faustov:2015eba}  & \cite{Santopinto:2014opa} & \cite{Menapara:2021dzi} &\cite{Oudichhya:2022off}  & \multicolumn{1}{c|}{\begin{tabular}[c]{@{}c@{}}Exp.\\ ~\cite{Workman:2022ynf} \end{tabular}}  \\ \hline \hline
$m(J^P=\frac{3}{2}^-)$ & 
$1513.64\pm 8.76$ & $1470.44\pm 16.08$& $1490$  &$1545$  & $1498$ & $1490$ & $1508$ & $1549$ & $1431$ &
 \multicolumn{1}{c|}{\begin{tabular}[c]{@{}c@{}}$1534$\\ $1544$\end{tabular}}  & $1551.23\pm0.43$ & $1519$   \\ \hline
$m'(J^P=\frac{3}{2}^-)$        &                                                                                                                                             $1687.91\pm 0.31$ & $1732.56 \pm 11.55$& $1690$ & $1645$ & $1629$ & $1690$ & $1662$ & $1693$ & $1650$ &  
  \multicolumn{1}{c|}{\begin{tabular}[c]{@{}c@{}}$1819$\\ $1841$\end{tabular}}  & - &$1690$\\ \hline
$\tilde{m}(J^P=\frac{3}{2}^+)$ &                                                                                                                                                                      $1882.37\pm11.95$ &$1843.47 \pm 11.21$&    -    & $1900$ & $1855$ &   -     & $1823$ & $1854$ & $1896$ & 
 \multicolumn{1}{c|}{\begin{tabular}[c]{@{}c@{}}$1769$\\ $1789$\end{tabular}}  & - & $1890$  \\ \hline
\end{tabular}
\caption{The mass results of the present work and that of different works and experimental masses for the $\Lambda$ states with spin-parity quantum numbers $J^P=\frac{3}{2}^{\pm}$ in units of MeV. }
\label{tab:differentworks}
\end{table}

As is seen, though there are various studies in literature with alternative methods about the $\Lambda$ baryon with spin $\frac{3}{2}$ such as $\Lambda(1520)$, $\Lambda(1690)$, and $\Lambda(1890)$, the discrepancies among the results necessitate more such works to provide further information over these particles. Studying the properties of these states helps us better identify these states, understand their nature, and interactions with other particles. These studies may contribute to the understanding of the underlying mechanisms of the interactions of these particles with others and their role in various particle physics processes. These investigations may help to gain information about the spin structure of these particles and the role of spin in their interactions with other particles and also be used to test the validity of various theoretical models in particle physics.  Besides, these studies deliver useful information that may be helpful to compare and comprehend the results of future experiments related to these baryons and their spin-parity quantum number assignments.

\section*{ACKNOWLEDGEMENTS}
 K. Azizi is grateful to Iran National Science Foundation (INSF) for the  partial financial support provided under the elites Grant No.  4025036.



\end{document}